**CROCODYLIA — CROCODILIANS**

***CAIMAN CROCODILUS*** (Spectacled Caiman)**. OPPORTUNISTIC FORAGING.** We document opportunistic foraging behavior by *Caiman crocodilus* in a post-inundation forest at Estacíon Biologíca Caño Palma, Costa Rica. Estacíon Biologíca Caño Palma is a 40-ha reserve located on the northeast coast of Costa Rica, south of Barra del Colorado. This reserve and the surrounding area is lowland tropical wet forest (*fide* Holdridge 1967. Lifezone Ecology, Tropical Science Center, San José, Costa Rica. 206 pp.) comprised predominantly of *Manicaria* swamp forest (Myers 1990. *In* Lugo et al. [eds.], Ecosystems of the World, pp. 267–278. Elsevier, Oxford, UK, 527 pp.). Bounded by large catchment rivers to the north, south, and west with a blackwater channel to the east, the forest inundates seasonally (November–December and May). *Manicaria* forests typically exhibit a prominent bi-annual polymodal inundation during the wettest season (Junk et al. 2000. The Central Amazon floodplain: Actual Use and Options for a Sustainable Management. Backhuys Publishers, Leiden, Holland. 584 pp.). The seasonal inundation event that engulfs Estacíon Biologíca Caño Palma is also coupled with local tidal flow patterns (Kelso 1967. A Contribution to the Ecology of a Tropical Estuary. M.Sc. Thesis, Univ. Florida, Gainesville. 156 pp.). Once inundation subsides, numerous temporal pools remain in the forest; these generally disappear during the warmer (and drier) months of the year.

In December 2002 and again in January 2004, during three months of weekly diurnal visual-encounter transects of post-inundation *Manicaria* forest, 5 adult and 7 juvenile caiman (2002) and 4 adult and 3 juvenile caiman (2004) respectively were located well into (often > 100 m) the forest. Once disturbed, they retreated terrestrially toward the channel associated with riparian habitat, rapidly slipping in and out of temporary pools as they headed in an easterly direction toward the main blackwater channel bordering the property.

Caiman reproduction, which generally occurs during the rainy season in this region of Costa Rica (November–February), involves construction of vegetation mounds in forested environments to incubate their eggs (Allsteadt 1994. J. Herpetol. 22:12–19.). As in other crocodilians, caiman exhibit well-developed parental care and will defend nests from predators, the primary threat to their eggs (Leenders 2001. A Guide to Amphibians and Reptiles of Costa Rica. Zona Tropical, Miami, Florida. 305 pp.). We detected no evidence of nests or nest-guarding behavior in the areas where we sighted caiman. However, these areas, still saturated by water, were found to frequently be full of suffocating fish trapped in desiccating flood pools or anurans utilizing these ponds for breeding. Fish species present included: *Archocentrus nigrofasciatus* and *Parachromis managuensis* (Cichlidae), *Rhamdia guatemalensis* and *Rhamdia rogersi* (Pimelodidae), and *Astyanax aeneus* (Characidae). *Atractosteus tropicus* (Order: Semionotiformes) were also hunting in these pools and may have constituted prey. That caiman were feeding on the diverse prey within these seasonally restricted environments seems likely. Fish are an important prey species for many crocodilians (Magnusson 1987. J. Herpetol. 21:85–95) and can make up over 25% of total prey items in sub-adult and mature adult caiman (Thorbjarnarson 1993. Herpetologica 49:108–117; Velasco et al. 1994. Crocodile Specialist Group Newsletter 13:20–21). In tropical blackwaters Cichlidae, Pimelodidae, and Characidae can make up 10–27% of the composition of fish species consumed (Thorbjarnarson 1993, *op. cit.*; Santos et al. 1996. Herpetol. J. 6:111–117) and possibly more when increases in water level like that which occurred along the Caño Palma allow more fish access to greater volumes of water and increased predation susceptibility (Silveira and Magnusson 1999. J. Herpetol. 33:181–192).

Caiman now occur in diverse habitats such as marshes, rivers, channels, and lakes in both the Caribbean and Pacific lowlands of Costa Rica, particularly as a result of the now-diminished ranges of sympatric competitors (*Crocodylus acutus*) (Magnusson 1982. Proc. 5th IUCN/SSC Croc. Spec. Group, pp. 108–116. Gland, Switzerland). They are commonly found in the canals, dikes, and channel networks in forested floodplain habitats (Allsteadt and Vaughan 1992. Brenesia 38:65–69; Guyer 1994. *In* McDade et al. [eds.], La Selva: Ecology and Natural History of a Neotropical Rain Forest, pp. 210–216. Univ. Chicago Press, Chicago, Illinois; Guyer and Donnelly 2005. Amphibians and Reptiles of La Selva, Costa Rica and the Caribbean Slope. Univ. California Press, Berkeley, California. 299 pp.; Ouboter and Nanhoe 1988. J. Herpetol. 22:283–294; Savage 2002. The Amphibians and Reptiles of Costa Rica: A Herpetofauna Between Two Continents, Between Two Seas. Univ. Chicago Press, Illinois. 934 pp.). This species is known to establish territories in local channel networks where they exhibit high site tenacity (Savage 2002, *op. cit.*), especially in areas that provide a sustainable food resource. Opportunistic seasonal shifts in habitat use has not been widely reported in *Caiman crocodilus*. Our observations imply that these habitat shifts provide enhanced feeding opportunities that might be unavailable the rest of the year. We present two models that might explain these habitat shifts. In the first, caiman might periodically abandon their territories within the permanent channel network during episodic inundation to specifically forage for trapped fish within temporary pools in the forest. The second model proposes that caiman, along with other fauna, advance into the forest during inundation following the expanding shoreline. As the water recedes, some aquatic animals are trapped in pools where they are vulnerable to amphibious predators, such as caiman. Caiman are able to escape back to traditional channel-margin habitat as the pools disappear. Which is correct provides an interesting question for further research.

We thank The Canadian Organization for Tropical Education and Rainforest Conservation for permission to study at Caño Palma Biological Station, Xavier Guevara of the Ministerio de Recursos Naturales Energia y Minas for permitting licenses, and Farnborough College of Science and Technology for assistance.

Submitted by **PAUL B. C. GRANT**, 4901 Cherry Tree Bend, Victoria B.C., V8Y 1SI, Canada; **TODD R. LEWIS**, Westfield, 4 Worgret Road, Wareham, Dorset, BH20 4PJ, United Kingdom (e-mail: biotropical@gawab.com); **THOMAS C. LADUKE**, East Stroudsburg University, 200 Prospect Street, East Stroudsburg, Pennsylvania 18301-2999, USA; and **COLIN RYALL**, Farnborough College of Technology, Farnborough, Hampshire, GU14 6SB. United Kingdom.